\documentclass[3p,times,twocolumn]{elsarticle}

\usepackage{ecrc}

\volume{00}

\firstpage{1}

\journalname{Nuclear and Particle Physics Proceedings}

\runauth{P.~Christiansen}

\jid{nppp}

\jnltitlelogo{Nuclear and Particle Physics Proceedings}

\usepackage{amssymb}

\usepackage[figuresright]{rotating}

\usepackage[english]{babel}
\usepackage[utf8]{inputenc}
\usepackage{amsmath}
\usepackage{amsfonts}
\usepackage{amssymb}
\usepackage{hyperref}
\usepackage{xspace}
\newcommand{\vtwo}{\ensuremath{v_{2}}\xspace}
\newcommand{\etwo}{\ensuremath{\varepsilon_2}\xspace}
\newcommand{\vthree}{\ensuremath{v_{3}}\xspace}
\newcommand{\Lin}{\ensuremath{L_{\rm in}}\xspace}
\newcommand{\Lout}{\ensuremath{L_{\rm out}}\xspace}
\newcommand{\pt}{\ensuremath{p_{\rm T}}\xspace}
\newcommand{\raa}{\ensuremath{R_{\rm AA}}\xspace}
\newcommand{\dndeta}{\ensuremath{dN/d\eta}\xspace}
\newcommand{\pp}{pp\xspace}

\newcommand{\pbpb}{Pb--Pb\xspace}
\newcommand{\gevc}[1]{\ensuremath{#1 \text{\,GeV/$c$}}\xspace}
\newcommand{\raain}{\ensuremath{R_{\rm AA,in}}\xspace}
\newcommand{\raaout}{\ensuremath{R_{\rm AA,out}}\xspace}
\newcommand{\snnt}[1]{\ensuremath{\sqrt{s_{\rm NN}} = #1 \text{\,TeV}}\xspace}
\newcommand{\snng}[1]{\ensuremath{\sqrt{s_{\rm NN}} = #1 \text{\,GeV}}\xspace}

\begin{document}

\begin{frontmatter}



\dochead{}

\title{High \pt spectra and anisotropy of light and heavy hadrons}


\author{Peter Christiansen}

\address{Lund University, Division of Particle Physics, Sweden}

\begin{abstract}
Data driven studies of heavy-ion results has played a big role in highlighting
interesting features of these complex systems. In this proceeding, a simple
QGP-brick interpretation of the \raa and \vtwo at high \pt ($\pt \approx
\gevc{10}$) is presented. This interpretation draws attention to two
fundamental questions: is there an effect of the asymmetric QGP expansion on
the \vtwo at high \pt? is there an energy loss difference between quarks and
gluons?  \\ Finally, it is discussed how these studies can be extended using
Event-Shape Engineering and how they can be applied to compare the energy loss
of light and heavy quarks.
\end{abstract}


\end{frontmatter}


\section{Introduction}
\label{sec:intro}

The understanding of heavy-ion data is a tremendous challenge. The heavy-ion
collision undergoes many phases from QGP creation to hadron production. Even
results obtained within models can be hard to interpret and approaches, such
as the core-corona model~\cite{Werner:2007bf}, has turned out to be powerful
in separating the physical origin of the observed centrality dependence of a
quantity (which typically has a simple explanation) from its absolute
magnitude (which typically requires the full QCD model).\\

In jet-quenching studies the challenge is one of the most formidable because
one has to model how the propagating jet interacts with the expanding bulk
medium. To be able to make a valid comparison of various jet quenching models
it has been important in the past to test them using a simple QGP brick (e.g.,
studies by the TECHQM Collaboration).

In this proceeding of Hard Probes 2016 (HP16) I report a study aimed at a
simple understanding of jet quenching in terms of essentially static QGP
bricks. The QGP bricks I will use are elliptic with semi-axes in (\Lin) and
out-of-plane (\Lout) given as the standard deviation of a standard Glauber
calculation. The area of the QGP cross section, $A$, is therefore $A =
4\pi\Lin\Lout$. The energy density, $\rho$, is assumed to be homogenous and
scale with \dndeta: $\rho = k\,\dndeta\,/\,A$, where $k$ is an unknown scale
factor that will be the same for all centralities and beam energies. All jets
are assumed to propagate from the center of the ellipse and so for each
centrality class there is basically two parameters the jet quenching can
depend on: path length $L$ and energy density $\rho$.

The studies described here have been published in
Ref.~\cite{Christiansen:2013hya} and extended somewhat in
Ref.~\cite{Christiansen:2016uaq}. At the HP16 conference several similar, but
much more advanced studies were presented by
J.~Noronha-Hostler~\cite{Noronha-Hostler:2016opk},
J.~Noronha~\cite{Prado:2016vgz}, and M.~Gyulassy, see also
Ref.~\cite{Noronha-Hostler:2016eow}.
 
\section{A scaling model for \raa and \vtwo}

\begin{figure*}[ht]
  \centering
  \includegraphics[keepaspectratio, width=0.75\columnwidth]{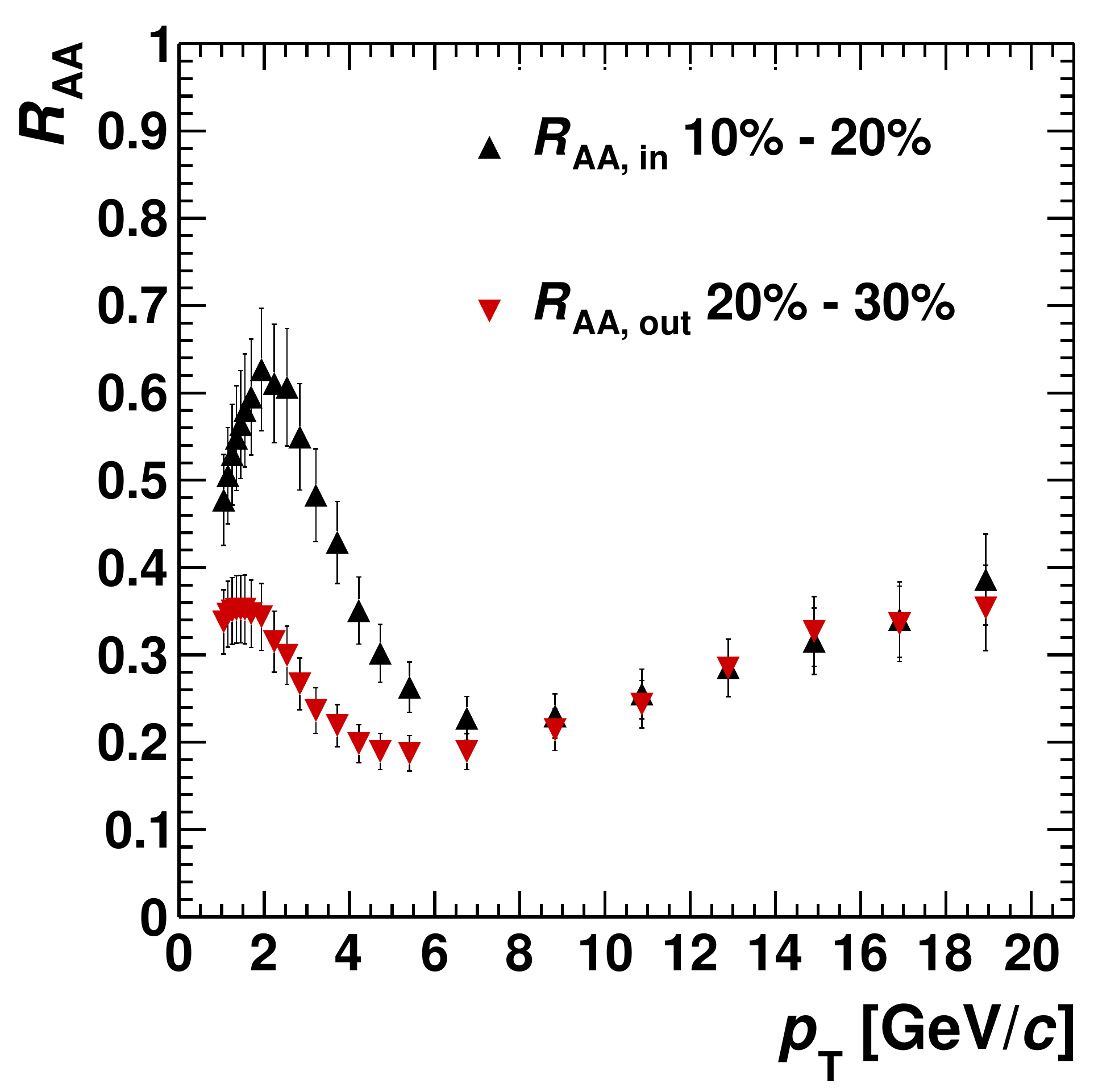}
  \includegraphics[keepaspectratio, width=0.75\columnwidth]{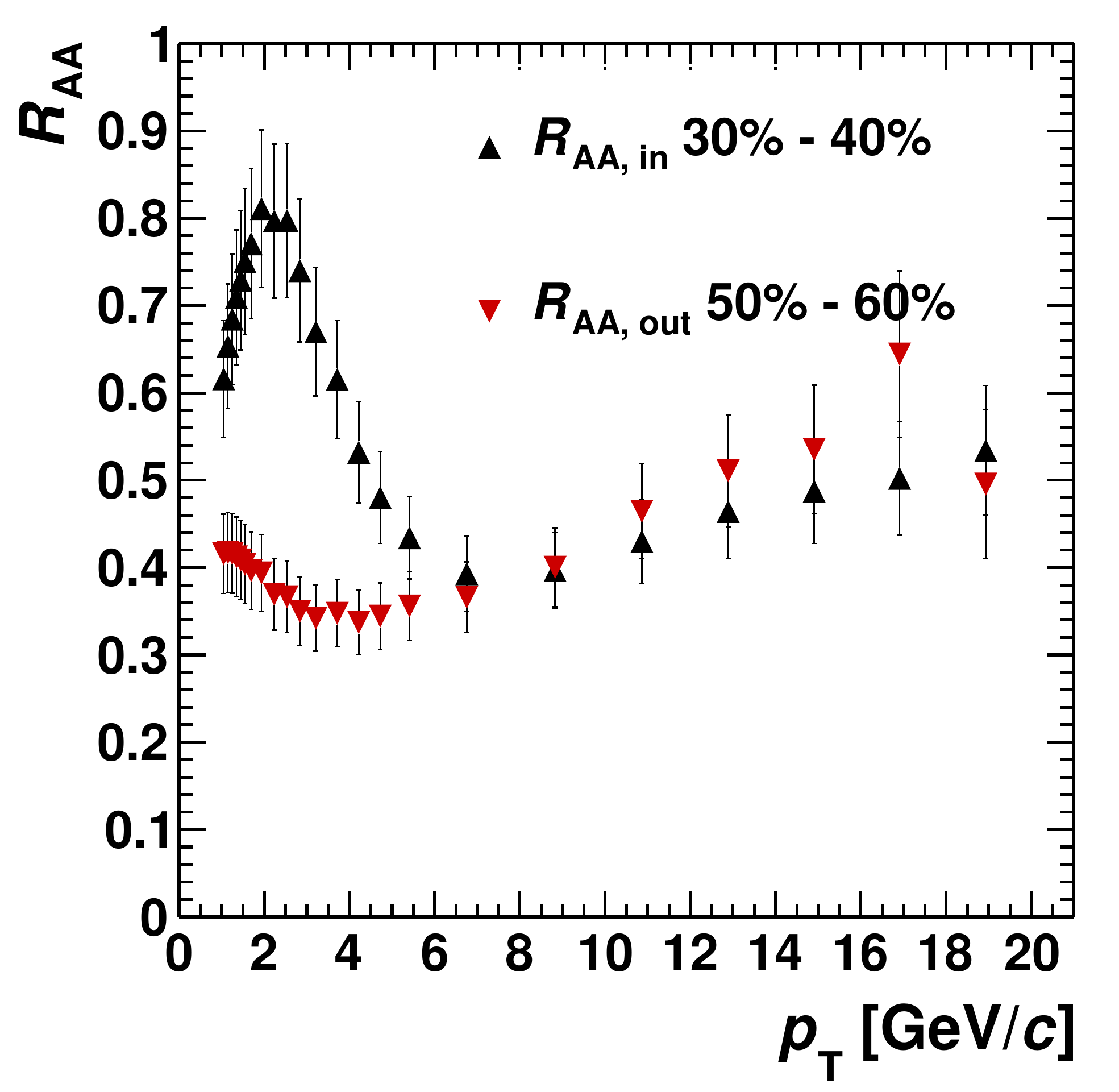}\\
  \includegraphics[keepaspectratio, width=0.75\columnwidth]{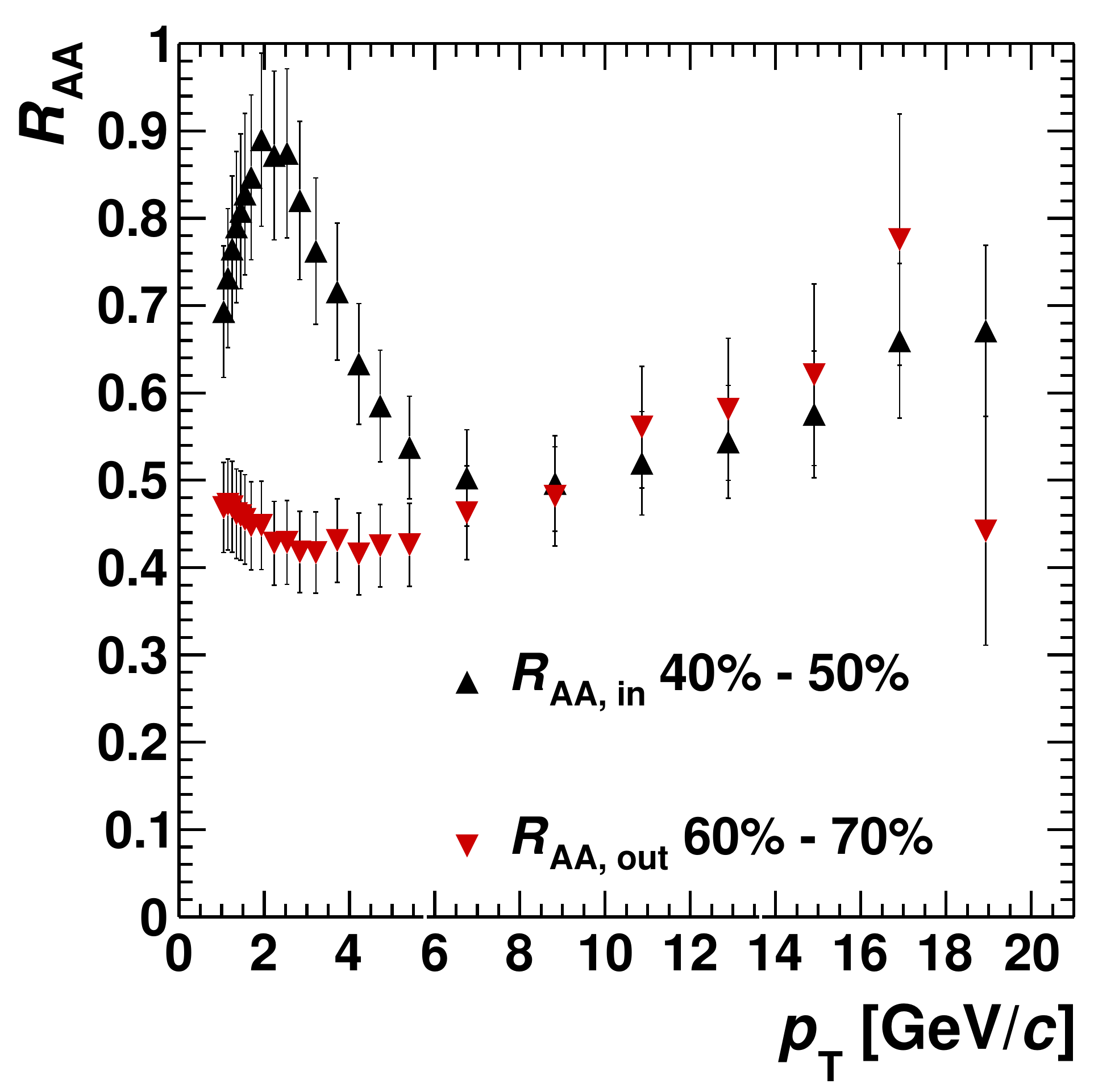}
  \includegraphics[keepaspectratio, width=0.75\columnwidth]{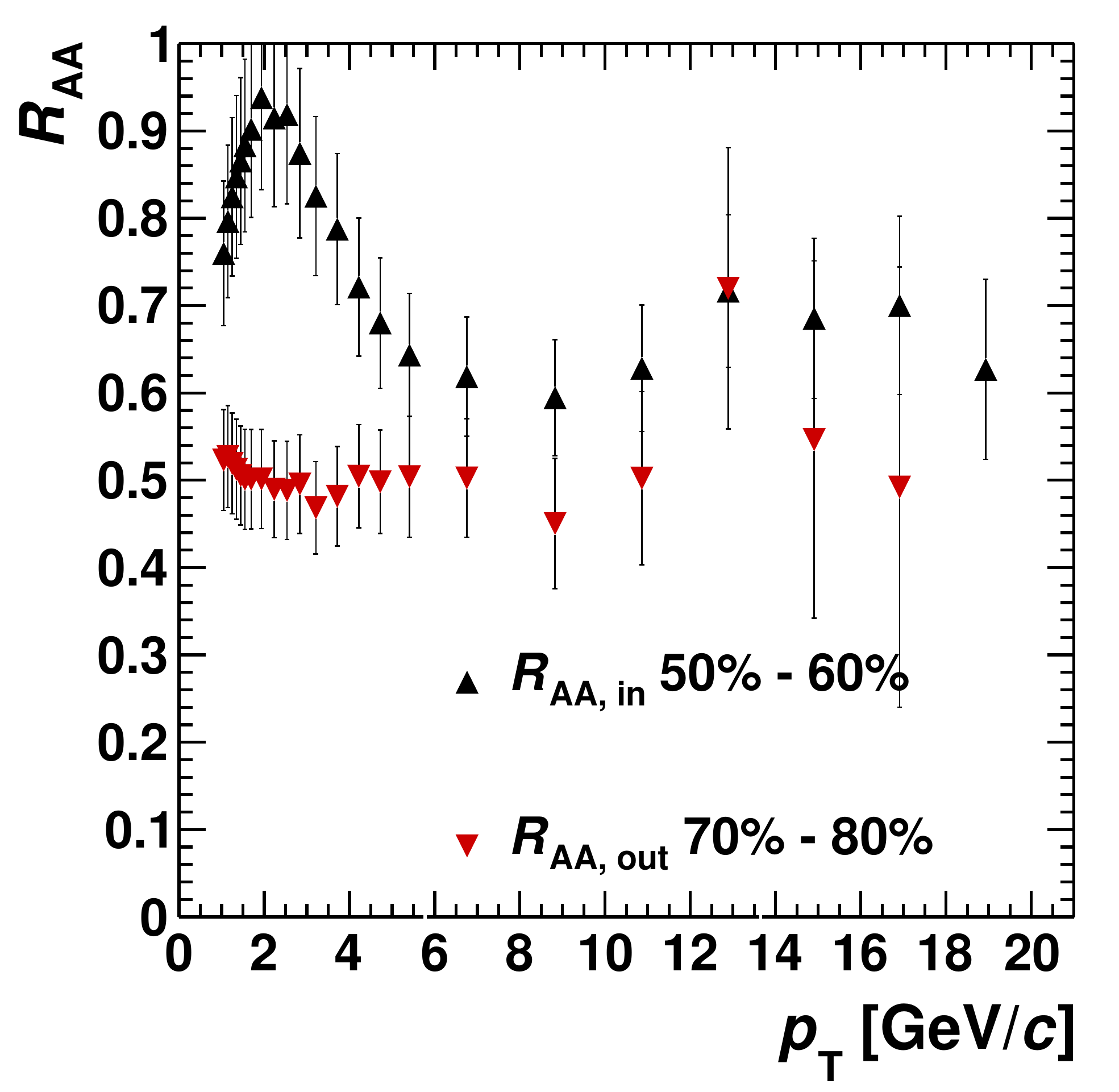}
  \caption{The comparison between \raa in- and out-of-plane for situations
    where the scaling variable $\sqrt{\rho} L$ is approximately the same. As
    can be seen, there is good agreement at high \pt where one expects jets to
    dominate while there is a large difference at lower \pt where the
    expansion in-plane is supposed to be stronger than out-of-plane.}
  \label{fig:raa_comparison}
\end{figure*}

The realization that many calculations could describe \raa but not \vtwo was a big
step forward in constraining jet quenching models,
e.g.~\cite{Adare:2010sp}. This lead to a lot of progress and recently
quite successful descriptions of \raa and \vtwo using Soft-Hard Event
Engineering~\cite{Noronha-Hostler:2016opk}.

In this proceeding the goal is much simpler. First, to describe the \raa and
\vtwo at $\pt \approx \gevc{10}$ for most centralities using a simple scaling
relation for the QGP brick parameters \Lin, \Lout, and $\rho$ (this
section). Second, to point out additional ways of trying to constrain jet
quenching models in data driven ways (the following two sections).

\subsection{High \pt tracks as proxies for jets}

One of the main results from jet studies at LHC is that quenched jets look
essentially like vacuum jets of the same final
energy~\cite{Chatrchyan:2012gw,Aad:2014wha}. While the CMS results are
consistent with no leading-\pt modification, the ATLAS results have smaller
systematic uncertainties and show some modification there. At HP16 it was
suggested by M.~Spousta that this could be due to the difference in quenching
of quark and gluon jets, which would lead to a different partonic composition of
quenched jets~\cite{Spousta:2015fca}. Here we assume that since the
modification is small, one can use high \pt tracks as proxies for jets.

At LHC there is evidence that a good definition of high \pt for jets is $\pt >
\gevc{10}$. At this \pt, the particle composition in \pbpb collisions is the
same as in \pp collisions~\cite{Adam:2015kca}. At the same time, the new
measurement by CMS of \vthree shows that it is consistent with zero for $\pt >
\gevc{10-20}$~\cite{CMS:2016uwf}. This suggests that for jet quenching only
the even harmonics are relevant and that for charged tracks all the relevant
information is contained in \raa and \vtwo.

\subsection{Constraining the path length}

The following is a brief summary of Ref.\cite{Christiansen:2013hya}.

Instead of describing \raa and \vtwo separately we will focus on describing
the \raa and \vtwo combined into the \raa in and out-of plane
\begin{eqnarray}
\label{eq:raain}  \raain(\pt) & \approx & (1 + 2 \vtwo(\pt))\raa(\pt)\\
\label{eq:raaout} \raaout(\pt) & \approx & (1 - 2 \vtwo(\pt))\raa(\pt),
\end{eqnarray}
using \snnt{2.76} results from ALICE~\cite{Abelev:2012hxa} and
ATLAS~\cite{ATLAS:2011ah}.

To attempt that, the pathlength dependence will be constrained using two
centrality classes where $\Lin \approx \Lout$. The question is then if we can
explain the difference between \raain and \raaout using only the difference in
energy density $\rho$ between the two centrality classes. I want to stress
here that this is not at all trivial because the expansion, which is not in
any way considered in this model, is very different in- and
out-of-plane. Furthermore, studies on the effect of this expansion on the
energy loss has previously found significant effects~\cite{Renk:2010qx}.

The scaling relation we finally obtain is that $\raain \approx \raaout$ when
they have the same scaling variable
$\sqrt{\rho}L$. Figure~\ref{fig:raa_comparison} shows that if we select
centrality classes based on this variable then indeed the \raain and \raaout
are the same. There is some tension for the most peripheral data, but this is
also the region where the origin of the large \vtwo is perhaps not completely
geometrical.\\

There are two important points to stress.

Since the eccentricity has a large centrality variation one would expect that
the asymmetric expansion, if important, would result in different scaling
relations. As this is not the case it raises the question if this asymmetric
expansion has any effect at all. This is a question that would be interesting
to study in realistic models. If one would find (an explanation for why) it
has no effects it would simplify the understanding of jet quenching
considerably.

\begin{figure}[htbp]
  \centering \includegraphics[keepaspectratio,
    width=0.95\columnwidth]{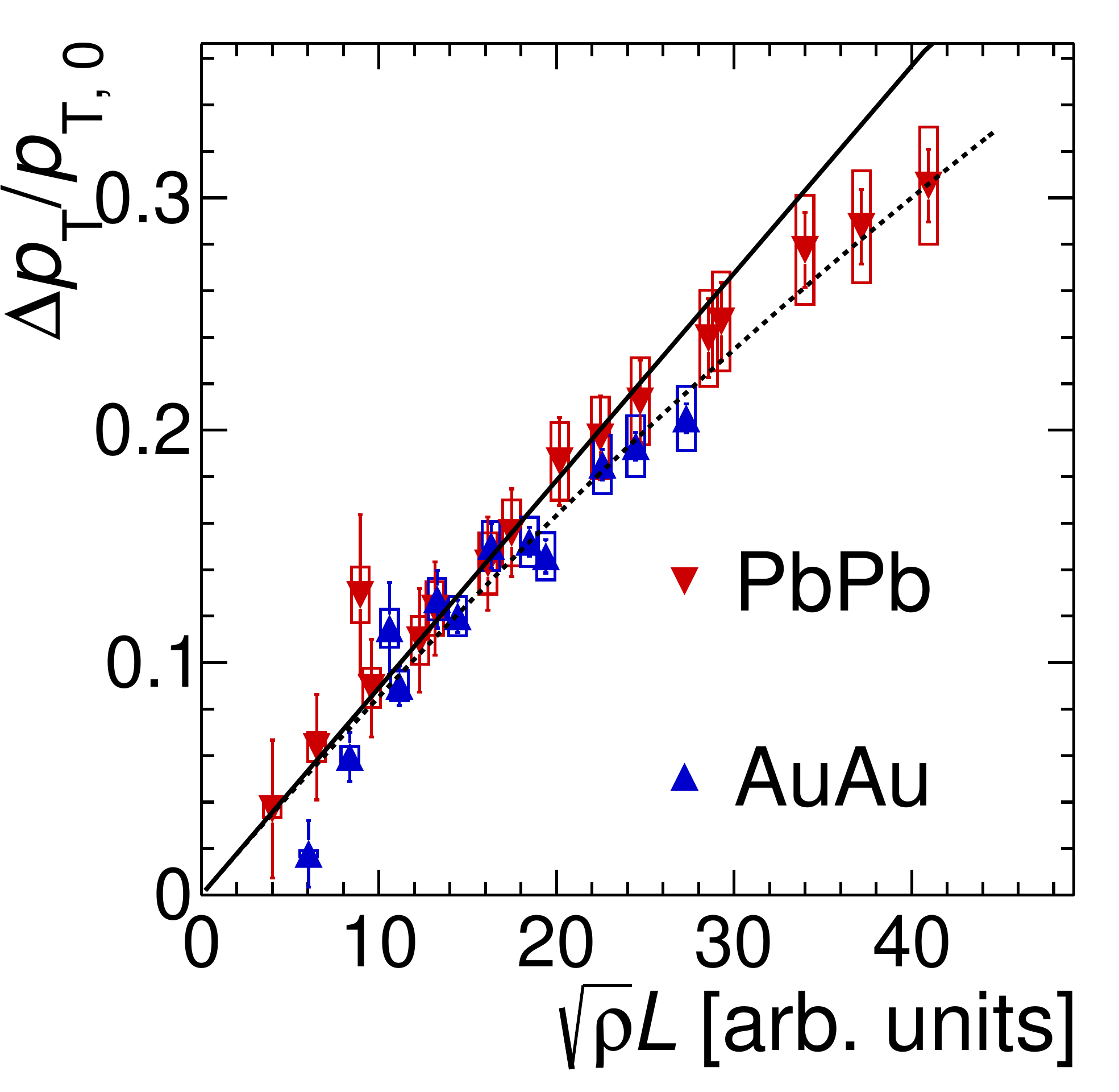}
  \caption{The \pt loss is found to scale with $\sqrt{\rho}L$ in exactly the
    same way at \snnt{2.76} (\pbpb), where the scaling relation was derived,
    and at \snng{200} (Au--Au, data taken from~\cite{Adare:2012wg}). The solid
    line is a linear fit. The dashed line is a fit that takes into account the
    decreasing \pt of the propagating particle leading to a small non-linear
    correction for large \pt losses.}
  \label{fig:pt_ratio}
\end{figure}

A scaling relation for \raain and \raaout is not unique because if
$\sqrt{\rho}L$ is a good scaling relation then so is $\rho L^2$ (and any power
of the scaling variable). What we did in Ref.~\cite{Christiansen:2013hya} was
to demand that the \pt loss was approximately linear in the scaling
variable. The \pt loss was derived using a similar estimate as employed by
PHENIX, see e.g.~\cite{Adare:2015cua}. Surprisingly, when we employed the same
scaling variable for PHENIX $\pi^0$ high-\pt results we found that they follow
the same trend as shown in Fig.~\ref{fig:pt_ratio}. The origin of this scaling
relation is very puzzling for us as the amount of quark and gluon jets as this
\pt is very different at these two energies.

\section{Event-Shape Engineering}

The following idea was also described in Ref.\cite{Christiansen:2016uaq}.

In the previous section the path length was constrained and the energy density
was varied. Using Event-Shape Engineering (ESE)~\cite{Schukraft:2012ah} it is
possible to constrain the energy density and vary the path length. The basic
idea is similar to the old idea of studying \raa and \vtwo together where one
also essentially does this. The new direction here is that one can obtain much
larger path length variations using ESE. ESE relies on the near perfect
fluidity of the QGP. This means that
\begin{equation}
\vtwo(\pt) = c(\pt) \etwo,
\end{equation}
where \etwo is the initial state ellipticity and $c(\pt)$ is a function
derivable from viscous hydrodynamic modeling of the QGP.

This means that, by selecting on the 2nd order flow vector of the event, one
can explore a much larger range of path lengths. This variation can be
estimated using Glauber calculations. For \pbpb 20--30\% central collisions
$\Lout \approx 1.31 \Lin$. Using ESE and selecting the 10\% of the events with
the highest \etwo one can obtain a variation of $\Lout \approx 1.62 \Lin$. At
the same time one is able to select events, e.g., the 10\% of the events with
the lowest \etwo, where the asymmetry is minimal in- and out-of-plane.

If one can use the variation of the initial state asymmetry to pin down the
effect of the radial expansion, then the additional path-length variation can
further constrain the path-length dependence.

\section{Energy loss of heavy quarks}

Many exciting results were shown for heavy quarks at HP16, e.g., B meson
\raa~\cite{CMS:2016jya} and D meson \vtwo~\cite{CMS:2016jtu}. There are
specific QCD predictions for what to expect for the relation between heavy
quarks and light quarks. To this author it seems attractive to make similar
data driven tests as has been done for light quarks and maybe some of the
additional ideas presented here could be included. At HP16 first model
calculations in this direction was presented~\cite{Prado:2016vgz}.

For reasons of statistics, the obvious first choice for comparing light and
heavy quark energy loss is the D meson. One problem is that the \pt region
used for the studies presented here ($\pt > \gevc{10}$) is probably too large
to detect a significant dead-cone effect for D mesons (c quarks), see
e.g.~\cite{Wicks:2005gt}. However, the results in
Fig.~\ref{fig:raa_comparison} suggests that if one would use \raaout for light
quarks then one could in principle go to lower $\pt \approx \gevc{4}$ without
having too big an effect of flow. Using ESE it might be possible to go even
lower in \pt. This could increase the sensitivity to the dead cone effect in
such studies~\footnote{In general flow effects should be smaller for pions
  than for protons and this could allow light quarks studies of jet quenching
  down to even lower \pt.}.

\section{Conclusions}

The work presented here has followed two basic ideas:
\begin{itemize}
\item Using elliptic flow to fix path length while varying the medium density
\item Using Event-Shape Engineering to fix the medium density while varying
  the path length
\end{itemize}
These approaches have been used here in a data driven way, but one could also
attempt to use them to interpret complex models. The data driven analysis
presented here pointed to some interesting questions:
\begin{itemize}
\item The role of the medium expansion for the energy loss
\item The difference in energy loss for quark and gluon jets (RHIC and LHC)
\end{itemize}
These are questions where the author hopes that clearer insights will be
available at the next HP conference.

\section*{Acknowledgements}

The author would like to thank the organizers for a wonderful conference:
amazing physics, fantastic food, and perfect liquid.

\nocite{*}
\bibliographystyle{elsarticle-num}
\bibliography{p_christiansen_hp_16}







\end{document}